\documentclass[preprintnumbers,article,amsmath,amssymb,floatfix,10pt,prd,onecolumn,
superscriptaddress,nofootinbib]{revtex4-2}
\usepackage{bm}
\usepackage{amsfonts}
\usepackage{latexsym}
\usepackage[latin1]{inputenc}
\usepackage{graphicx}
\usepackage{amsmath}
\usepackage{palatino}
\usepackage{mathpazo}
\usepackage{textcomp}
\usepackage{float}
\usepackage{booktabs}
\usepackage{dcolumn}
\usepackage{ragged2e}
\usepackage{hyperref}
\hypersetup{colorlinks,citecolor=blue}
\hypersetup{colorlinks=true,linkcolor=red,filecolor=magenta,    urlcolor=blue}
\usepackage{amsmath}
\usepackage{xcolor}
\usepackage{orcidlink}
\usepackage{epsfig}
\usepackage{subfigure}
\usepackage{commath}

\def\jnl@style{\it}
\def\aaref@jnl#1{{\jnl@style#1}}

\def\aaref@jnl#1{{\jnl@style#1}}

\def\aj{\aaref@jnl{AJ}}                   
\def\apj{\aaref@jnl{ApJ}}                 
\def\apjl{\aaref@jnl{ApJ}}                
\def\apjs{\aaref@jnl{ApJS}}               
\def\apss{\aaref@jnl{Ap\&SS}}             
\def\aap{\aaref@jnl{A\&A}}                
\def\aapr{\aaref@jnl{A\&A~Rev.}}          
\def\aaps{\aaref@jnl{A\&AS}}              
\def\mnras{\aaref@jnl{Mon.~Not.~Roy.~Astron.~Soc.}}             
\def\prd{\aaref@jnl{Phys.~Rev.~D}}        
\def\prc{\aaref@jnl{Phys.~Rev.~C}}  
\def\prl{\aaref@jnl{Phys.~Rev.~Lett.}}    
\def\qjras{\aaref@jnl{QJRAS}}             
\def\skytel{\aaref@jnl{S\&T}}             
\def\ssr{\aaref@jnl{Space~Sci.~Rev.}}     
\def\zap{\aaref@jnl{ZAp}}                 
\def\nat{\aaref@jnl{Nature}}              
\def\aplett{\aaref@jnl{Astrophys.~Lett.}} 
\def\apspr{\aaref@jnl{Astrophys.~Space~Phys.~Res.}} 
\def\physrep{\aaref@jnl{Phys.~Rep.}}      
\def\physscr{\aaref@jnl{Phys.~Scr}}       
\def\commat{\aaref@jnl{Comm.~Math.~Phys.}}              
\def\science{\aaref@jnl{Science}}               
\def\cqg{\aaref@jnl{Classical Quant.~Grav.}}            
\def\jpcs{\aaref@jnl{JPCS}}                                     
\def\ijmpd{\aaref@jnl{Int.~J.~Mod.~Phys.~D}}                    
\def\grg{\aaref@jnl{Gen.~Relat.~Gravit.}}               
\def\rpp{\aaref@jnl{Rep.~Prog.~Phys.}}          
\def\npa{\aaref@jnl{Nucl.~Phys.~A}}        
\def\lrr{\aaref@jnl{Living Rev.~Rel.}}                   
\def\jcap{\aaref@jnl{J.~Cosmology Astropart.~Phys.}}    
\def\rmp{\aaref@jnl{Rev.~Mod.~Phys.}}   
\def\epjc{\aaref@jnl{Eur.~Phys.~J.~C}} 
\def\plb{\aaref@jnl{~Phy.~Lett.~B}} 
\def\mpla{\aaref@jnl{Mod.~Phy.~Lett.~A}} 
\def\arxiv{\aaref@jnl{arxiv.org}}

\allowdisplaybreaks[1]

\begin{document}

\color{black}       

\title{Bouncing cosmological models in $f(R,L_m)$ gravity}

\author{Lakhan V. Jaybhaye\orcidlink{0000-0003-1497-276X}}
\email{lakhanjaybhaye@gmail.com}
\affiliation{Department of Mathematics, Birla Institute of Technology and
Science-Pilani,\\ Hyderabad Campus, Hyderabad-500078, India.}

\author{Raja Solanki\orcidlink{0000-0001-8849-7688}}
\email{rajasolanki8268@gmail.com}
\affiliation{Department of Mathematics, Birla Institute of Technology and
Science-Pilani,\\ Hyderabad Campus, Hyderabad-500078, India.}

\author{P. K. Sahoo\orcidlink{0000-0003-2130-8832}}
\email{pksahoo@hyderabad.bits-pilani.ac.in}
\affiliation{Department of Mathematics, Birla Institute of Technology and
Science-Pilani,\\ Hyderabad Campus, Hyderabad-500078, India.}
\affiliation{Faculty of Mathematics \& Computer Science, Transilvania University of Brasov, Eroilor 29, Brasov, Romania}

\date{\today}

\begin{abstract}

This article explores matter bounce non-singular cosmology in $f(R,L_m)$ gravity. We consider two non-linear $f(R,L_m)$ functional forms, specifically, $f(R,L_m) = \frac{R}{2} + \lambda R^2 + \alpha L_m$  and $f(R,L_m) = \frac{R}{2} + L_m ^\beta + \gamma$ representing a minimal coupling case. We derive the corresponding Friedmann-like equations for both the assumed models in the FLRW background, and then we present the impact of the model parameters along with the parameter of bouncing scale factor on the equation of state parameter, pressure, and the energy density. In addition, we examine the dynamical behavior of cosmographic parameters such as jerk, lerk, and snap parameters. Further, we find that the violation of the null energy condition along with the strong energy condition depicts the non-singular accelerating behavior, corresponding to both assumed non-linear $f(R,L_m)$ functions. Lastly, we present the behavior of the adiabatic speed of sound to examine the viability of the considered cosmological bouncing scenario. 
\end{abstract}

\maketitle

\section{Introduction}\label{sec1}
\justify
The field of cosmology underwent a significant transformation following the confirmation of the accelerating nature of cosmic expansion through observational evidence obtained from searches for type Ia supernovae \cite{Riess,Perlmutter}. Over the past twenty years, numerous observational findings, including those from the Wilkinson Microwave Anisotropy Probe \cite{D.N.},  Baryonic Acoustic Oscillations \cite{D.J.,W.J.}, Cosmic Microwave Background Radiation \cite{C.L.,R.R.}, and the Large Scale Structure \cite{T.Koivisto,S.F.}, have supported this accelerating phenomenon. The conventional understanding of cosmology has strongly endorsed dark energy models to describe this observed acceleration. The most notable explanation for dark energy is the cosmological constant $\Lambda$, which arises due to vacuum quantum energy \cite{S.W.}. Despite fitting well with observational data, the cosmological constant $\Lambda$ is plagued by two significant issues, the cosmological constant problem and the coincidence problem. The value obtained for it from particle physics exhibits a high discrepancy of almost 120 orders of magnitude when compared to the value suggested by several cosmological observations \cite{E.J.}. 

An alternative approach to avoid the need for undetected dark energy involves considering a more general description of the gravitational field. This can be achieved through the use of cosmological models with modified action of general relativity. Theoretical models that replace the standard action of the general relativity with an arbitrary functional form $f(R)$ of the Ricci scalar $R$, were introduced in \cite{H.A.,R.K.}. References \cite{Shin,Sal,Alex} present observational signatures of cosmological models in $f(R)$ gravity, as well as constraints imposed by the solar system and equivalence principle. Viable $f(R)$ gravity models have been established through solar system tests \cite{Noj,V.F.,L.A.}. References \cite{Noj-2,Noj-3,G.C.,J.S.,R.C.} offer insights into the several applications of models in $f(R)$ gravity. In a study, a modified version of $f(R)$ gravity that included a direct coupling of the function $f(R)$ with the Lagrangian density of matter was introduced \cite{O.B.}. The inclusion of non-minimal matter curvature couplings has essential implications for astrophysics and cosmology. Harko investigated the effect of curvature and matter coupling in Weyl geometry on galactic rotation curves, thermodynamical features, the matter Lagrangian, and the energy-momentum tensor using non-minimal couplings \cite{THK-2,THK-3,THK-4,THK-5}. Bertolami et al. \cite{THK-EX} and Faraoni \cite{V.F.-2}, have also explored the possibility of matter-curvature couplings in extended gravity and established the feasibility criterion with an extra force. Further, Harko and Lobo introduced an extension of matter curvature coupling theories in the form of an $f(R,L_m)$ gravity theory, where $f(R,L_m)$ is a generic function that depends on the curvature term $R$ and the matter Lagrangian $L_m$ \cite{THK-6}. The models of $f(R,L_m)$ gravity theory are constrained by solar system tests \cite{FR,JP}. Several new exciting results on $f(R,L_m)$ gravity have been reported recently \cite{APP,RV-1,RV-2,JWK,Jayb,Jayb1,Jayb2}.

It is well known that the inflationary scenario proposes that the Universe underwent rapid expansion at its beginning. However, the singularity occurred before inflation began, so the scenario needs to be completed to explain the entire history of the Universe. An alternate route to resolve this issue is to consider the matter bounce scenario \cite{MB-1,MB-2,Haro,Saho}, which suggests that the Universe first contracted before expanding without encountering a singularity. This theory proposes an initial matter-dominated contraction epoch, followed by a bounce that further causes a generation of casual fluctuations. However, it should be noted that in a flat Universe, the non-singular bounce may violate the null energy condition. Such non-singular cosmology with the violation of the null energy condition is supported in generalized Galileon theories \cite{MB-3}. The idea of a big bounce scenario in the context of modified gravity replacing the big bang singularity is an intriguing topic for research \cite{Mand2,Zuba,Bhat,Yous}. There are plenty of interesting results of bouncing scenario in the context of $f(R)$ gravity that have appeared in the literature, such as Carlos et al. \cite{Barr} have studied bouncing cosmologies in Palatini $f(R)$ gravity formalism, Niladri et al. \cite{Paul} studied cosmological bounce in spatially flat FRW environment in metric $f(R)$ gravity, S. D. Odintsov and V. K. Oikonomou investigated matter bounce loop quantum cosmology from $f(R)$ gravity \cite{Odin}, R. Myrzakulov and L. Sebastian studied bouncing solutions with viscous fluid \cite{Myrz}, bouncing cosmology with future singularity from modified gravity has been investigated in \cite{Odin1}, and others \cite{Ilya,Aman,Odin2,sing,gadb,sing1,bane,caru}.
The present work investigates matter bounce non-singular cosmological scenario in the $f(R,L_m)$ gravity background. In sec. \ref{sec2}, we present the mathematical framework of $f(R,L_m)$ gravity. In sec. \ref{sec3}, we present the idea of matter bounce. We specify a scale factor parametrization and investigate the intricate dynamical development of a universe undergoing a non-singular bounce. Further, we discuss the cosmography of the bouncing model in terms of cosmic time in this section. In sec. \ref{sec5} we investigate matter bounce scenario in $f(R,L_m)$ gravity framework using two non-linear $f(R,L_m)$ function. Further in sec. \ref{sec5}, we derive the Friedmann like equations and the corresponding dynamical parameters $\rho$, $p$, and $\omega$ corresponding to assumed non-linear $f(R,L_m)$ functional forms. Then, we have presented the profiles of different energy conditions. Moreover, we investigate the stability of the considered bouncing scenarios. Finally in Sec.\ref{sec6}, we presented outcomes of our findings.

\section{Motion equations in $f(R,L_m)$ Gravity}\label{sec2}
\justifying

\justify The geometrically modified action with the matter for $f(R,L_m)$ gravity formulation is given by \cite{THK-6}

\begin{equation}\label{2a}
S= \int{f(R,L_m)\sqrt{-g}d^4x}.
\end{equation}

\justify In which $f(R,L_m)$ is an arbitrarily chosen function of $R$ (Ricci scalar) and $L_m$ (matter Langragian density). The energy momentum tensor $T_{\mu \nu}$, of the matter is defined as

 \begin{equation}\label{2b}
T_{\mu\nu} = \frac{-2}{\sqrt{-g}} \frac{\delta(\sqrt{-g}L_m)}{\delta g^{\mu\nu}}.
\end{equation}

\justify The field equation is obtained by varying the action \eqref{2a} concerning the metric tensor $g^{\mu\nu}$ as
 
\begin{equation}\label{2c}
f_R R_{\mu\nu} + (g_{\mu\nu} \square - \nabla_\mu \nabla_\nu)f_R - \frac{1}{2} (f-f_{L_m}L_m)g_{\mu\nu} = \frac{1}{2} f_{L_m} T_{\mu\nu},
\end{equation}

\justify with $\square=g^{\mu\nu}\nabla_\mu \nabla_\nu$, $f_R= \frac{\partial f}{\partial R}$, and $f_{L_m}=\frac{\partial f}{\partial L_m}$. Then the contraction of equation \eqref{2c} offers

\begin{equation}\label{2d}
R f_R + 3\square f_R - 2(f-f_{L_m}L_m) = \frac{1}{2} f_{L_m} T.
\end{equation}

\justify Moreover, one can acquire the following result by taking covariant derivative in equation \eqref{2c}
\begin{equation}\label{N11}
\nabla^\mu T_{\mu\nu} = 2\nabla^\mu ln(f_{L_m}) \frac{\partial L_m}{\partial g^{\mu\nu}}.
\end{equation}

\justify Where $T$ is the trace of energy-momentum tensor. Our Universe is homogeneous and isotropic on a large scale, according to recent CMB data. As a result, in the analysis presented here, we consider a flat FLRW background geometry in Cartesian coordinates with a metric,

\begin{equation}\label{2e}
ds^2= -dt^2 + a^2(t)\big(dx_1 ^2+dx_2 ^2+dx_3 ^2\big).
\end{equation}

\justify Where, $ a(t) $ is the scale factor of Universe. Furthermore, the Ricci scalar generated by the metric \eqref{2d} is as follows:

\begin{equation}\label{2f}
R= 6 ( \dot{H}+2H^2 ).
\end{equation}

\justify The stress-energy tensor describing the matter-energy distribution of the universe having perfect fluid, for the line element \eqref{2e} reads as,

\begin{equation}\label{n1}
T_{\mu\nu}=(\rho+p)u_\mu u_\nu+pg_{\mu\nu}.
\end{equation}

\justify Here $\rho$ and $p$ respectively denotes the usual matter density and isotropic pressure, with $u_\mu = (1, 0, 0, 0)$ as the four velocities vector. The corresponding Friedmann equations for the metric \eqref{2e} and arbitrary $f(R, L_m)$ function given as

\begin{equation}\label{n2}
\frac{6H^2 f_R+(f-Rf_R-L_m f_{L_m} )+6H\dot{f_R}}{f_{L_m}}=\rho,
\end{equation}

\begin{equation}\label{n3}
\frac{2\dot{H}f_R+6H^2 f_R-(f-L_m f_{L_m} )-4H\dot{f_R}-2\ddot{f_R}}{f_{L_m}}=p.
\end{equation}

\section{The Bouncing Model}\label{sec3}
\justifying
\justify To understand the behaviour of the Universe, we must first understand its physical and dynamical entities. The Hubble parameter is used to assert all physical parameters. It is worth noting that its inflationary scenario does not account for the entire previous history of the cosmos. As a result, the matter bounce has been suggested as a potential solution to this problem. We are also focused on the matter bounce scenario with the respective scale factor in this work as \cite{Zuba}

\begin{equation}\label{3a}
a(t)=(a_0 ^2 +\zeta^2 t^2)^\frac{1}{2}.
\end{equation}

\begin{figure}[H]
\centering
\subfigure[]{\includegraphics[width=7.5cm,height=5cm]{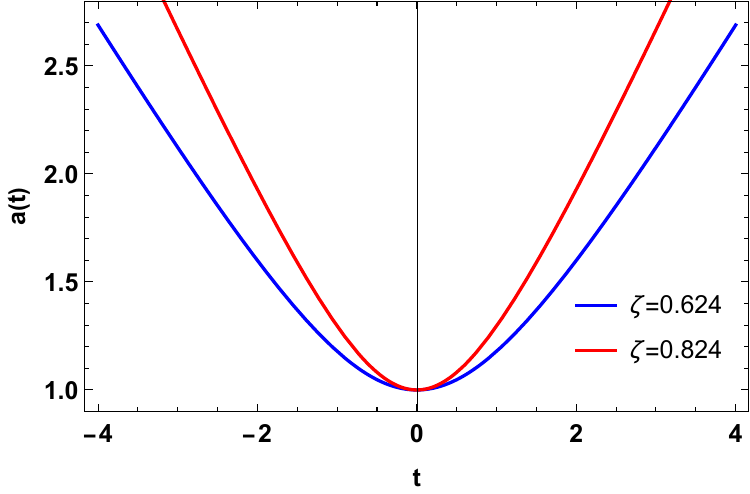}}\,\,\,\,\,\,\,\,\,\,\,\,
\subfigure[]{\includegraphics[width=7.5cm,height=5cm]{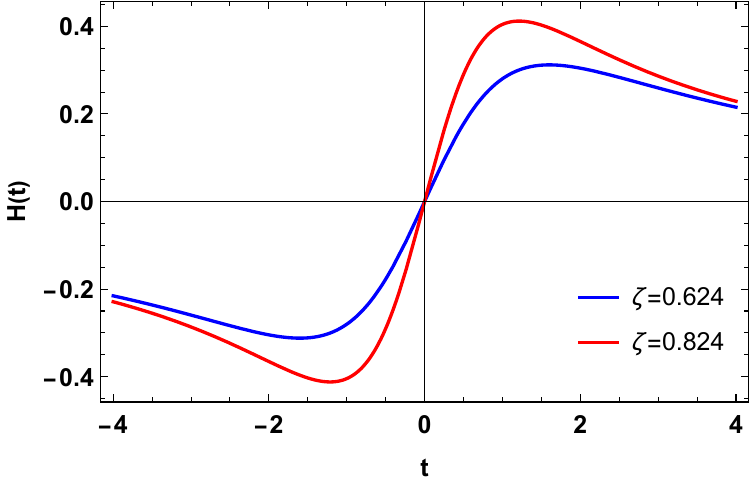}}\,\,\,\,\,\,\,\,\,\,\,\,
\subfigure[]{\includegraphics[width=7.5cm,height=5cm]{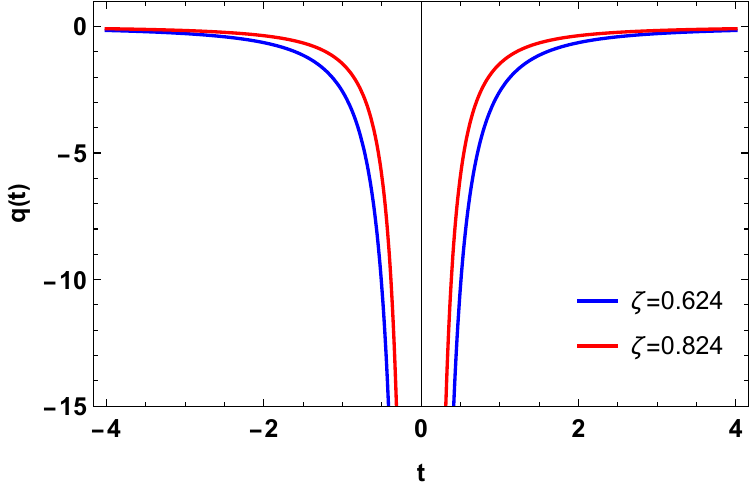}}
\caption{Profile of scale factor $a(t)$, Hubble parameter $H(t)$, and deceleration parameter $q(t)$ for $a_0=1$, with different values of $\zeta$ against cosmic time.}\label{1}
\end{figure}

\justify As this study is for the accelerated expansion phase, $\zeta$ is a positive bouncing parameter. The radius of the scale factor at the bouncing point $t=0$, is denoted by $a_0$. Figure \ref{1} (a) depicts the behaviour of the scale factor versus cosmic time for various values of $\zeta$. The scale factor expands symmetrically on both sides of the bounce point $t=0$, implying an initial contraction, a bounce, and then an expansion. The bouncing parameter $\zeta$ influences the slope of scale factor. In summary, bouncing parameter is an essential component that affects the slope of a curve of the scale factor $a$. The corresponding expression for the Hubble parameter of the scale factor \eqref{3a} is defined as 

\begin{equation}\label{3b}
H(t)=\frac{\zeta^2 t }{a_0 ^2  + \zeta^2 t^2}.
\end{equation}

\justify The Hubble parameter has negative values before the bounce as well as positive values after the bounce, which disappear at the bounce. Figure \ref{1} (b) depicts the adaptive behaviour of the Hubble parameter in the preceding bouncing scenario. The Hubble parameter meets the requirements for bouncing cosmology because it starts out negative, passes through zero at t = 0, and then turns positive. The deceleration parameter for the bouncing scenario described above is given by

\begin{equation}\label{3c}
q(t)=-\frac{a_0 ^2}{\zeta^2 t^2 }.
\end{equation}

\justify As a result, Figure \ref{1} (c) depicts the behavioral patterns of the deceleration parameter, whose negative range indicates an accelerated phase of expansion. The deceleration parameter exhibits a singularity near the bounce due to the bouncing model's selected responses. With this favourable behaviour of the basic parameters, we can now proceed with the chosen scale factor to evaluate the cosmological model.

\justify The traditional cosmographic approach is based on the Taylor series expansion of material objects. The cosmography is a great way to bridge the gap between cosmological models and popular approaches to understanding the dynamics of the universe \cite{Viss,Alam,Sahn,Capo1,Capo2}. The cosmographic set comprises cosmological parameters such as the Hubble parameter ($H$), deceleration parameter($q$), jerk parameter ($j$), snap parameter ($s$), and lerk parameter($l$) \cite{Avil,Sale,Mand}. The first two Hubble and deceleration parameters are addressed above in the context of FRW cosmology, while the other cosmographic parameters are obtained as follows: 

\begin{equation}\label{5a}
j(t)= -\frac{3a_0 ^2}{\zeta^2 t^2},
\end{equation}

\begin{equation}\label{5b}
s(t)=-\frac{3 \left(a_0^4 -4 a_0^2 \zeta ^2 t^2\right)}{\zeta ^4 t^4},
\end{equation}

\begin{equation}\label{5c}
l(t)=\frac{15 \left(3a_0^4-4 a_0^2 \zeta ^2 t^2\right)}{\zeta ^4 t^4}.
\end{equation}

\begin{figure}[H]
\centering
\subfigure[]{\includegraphics[width=7.5cm,height=5cm]{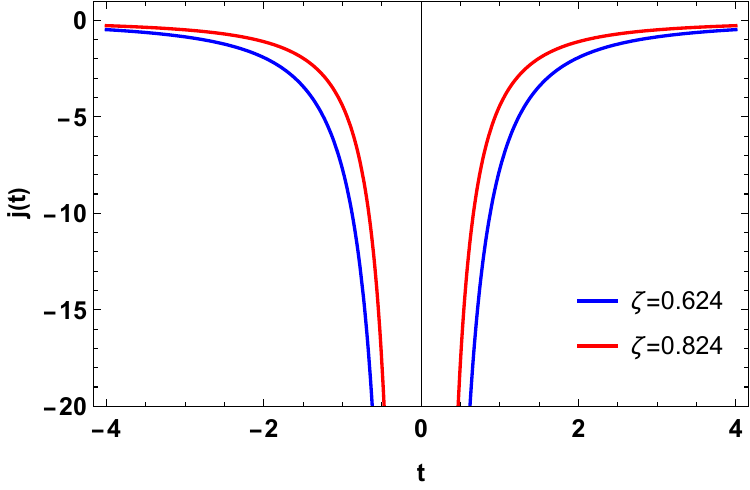}}\,\,\,\,\,\,\,\,\,\,\,\,
\subfigure[]{\includegraphics[width=7.5cm,height=5cm]{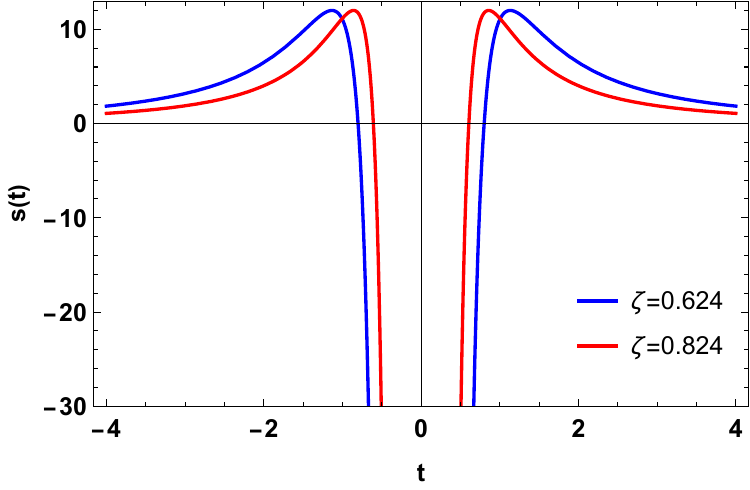}}\,\,\,\,\,\,\,\,\,\,\,\,
\subfigure[]{\includegraphics[width=7.5cm,height=5cm]{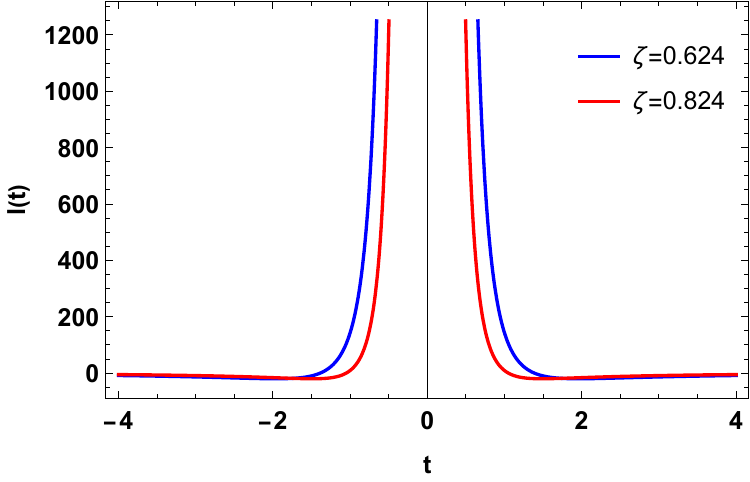}}
\caption{Profile of jerk parameter, snap parameter, and lerk parameter for $a_0=1$, with different values of $\zeta$ against cosmic time.}\label{2}
\end{figure}

\justify Figure \ref{2} (a) depicts the graphical representation of jerk parameter behaviour for the chosen parameters, whereas Figure \ref{2} (b) depicts the snap parameter behaviour. Regardless of the negative and positive time zone, the $\Lambda$CDM behaviour leads to the jerk parameter, which displays a singular bounce during the bouncing epoch. Similar patterns of behaviour are displayed by the snap parameter, however it deviates from the intended $\Lambda$CDM behaviour. But for the temporal area and the progression of universe is away from a matter-dominated one, the lerk parameter remains positive which is shown in Figure \ref{2} (c).

\section{Matter bounce scenario in $f(R,L_m)$ Gravity}\label{sec5}
\justifying
\justify In order to probe the implications of the assumed bouncing model under the $f(R,L_m)$ gravity background, we consider the following two non-linear $f(R,L_m)$ functional form, \\

\justify \textbf{Model I :} We consider the model $f(R,L_m) = \frac{R}{2}+ L_m ^{\beta}+\gamma$. The assumed minimal coupling model is motivated by an interesting work of Bose et al. \cite{Bose} in the $f(R, T)$ gravity. The assumed model $f(R,L_m)$ reduces to the $\Lambda$CDM case for the parameter values $\beta=1$, whereas it reduces to the GR case for the values $\beta=1$ and $\gamma=0$. The corresponding modified Friedmann equations with $L_m=\rho$ and for a perfect fluid distribution reads as

\begin{equation}\label{n4}
3H^2 = (2\beta-1) \rho^{\beta}-\gamma,
\end{equation}

\begin{equation}\label{n5}
-2\dot{H}-3H^2 =\big( \beta(p-\rho)+\rho\big) \rho^{\beta-1}+\gamma.
\end{equation} 

\justify Moreover, the equation \eqref{N11} yields the following energy-balance equation corresponding to the model I,

\begin{equation}\label{cc1}
 (2\beta-1)\dot{\rho}+ 3H(\rho+p)=0.   
\end{equation}

\justify \textbf{Model II :} We consider the model $f(R,L_m) = \frac{R}{2}+ \lambda R^2 + \alpha L_m $. The assumed model $f(R,L_m)$ reduces to the  GR case for the values $\lambda=0$ and $\alpha=1$. We choose this specific model due to its significance within contemporary theoretical physics, especially within the domain of modified gravity theories and cosmology. This functional form draws inspiration from the Starobinsky model, which has been thoroughly explored and confirmed to be consistent across various cosmological scenarios, including inflation \cite{Chai}, post-inflationary gravitational wave production \cite{Kosh}, and the evolution of star clusters \cite{Manz}. The corresponding modified Friedmann equations with $L_m=\rho$ and for a perfect fluid distribution reads as
 
\begin{equation}\label{2g}
3H^2 = \frac{\alpha \rho - 36\lambda(2 H \ddot{H}-\dot{H}^2)}{1+72\lambda\dot{H}},
\end{equation}

\begin{equation}\label{2h}
-2\dot{H}-3H^2 =\alpha p+12\lambda(2\dddot{H}+9\dot{H}^2+18\dot{H}H^2+12\ddot{H}H).
\end{equation} 

\justify Moreover, the equation \eqref{N11} yields the following energy-balance equation corresponding to the model II,

\begin{equation}\label{cc2}
 \dot{\rho}+ 3H(\rho+p)=0.   
\end{equation}

\subsection{Dynamical parameters}\label{sbsec1}
\justifying
\justify It is essential to look into the model's decisive parameters to comprehend the effective elements that make up the Universe. The action of decisive parameters indicates whether or not the model is bouncing. The decisive parameters such as energy density $\rho$, pressure $p$ and the EoS parameter $\omega$ for the $f(R,L_m)$ model I, obtained by using equation \eqref{3b} in \eqref{n4} and \eqref{n5} are as follows:

\begin{equation}\label{n6}
\rho(t)=\left(\frac{\gamma+\frac{3\zeta^4 t^2 }{(a_0^2+\zeta^2 t^2)^2}}{(2\beta-1)}\right)^\frac{1}{\beta},
\end{equation}

\begin{equation}\label{n7}
p(t)=-\frac{\left(a_0^4 \beta  \gamma +2 a_0^2 \zeta ^2 \left(\beta  \left(\gamma  t^2+2\right)-1\right)+\zeta ^4 t^2 \left(\beta  \left(\gamma  t^2-1\right)+2\right)\right)\left(\gamma+\frac{3\zeta^4 t^2 }{(a_0^2+\zeta^2 t^2)^2}\right)^\frac{1}{\beta}}{\beta (2\beta-1)^\frac{1}{\beta}\big(a_0 ^4 \gamma +2a_0^2 \gamma \zeta ^2 t^2 + (3+\gamma t^2)\zeta^4t^2\big)},
\end{equation}

\begin{equation}\label{n8}
\omega(t)=-\frac{a_0^4 \beta  \gamma +2 a_0^2 \zeta ^2 \left(\beta  \left(\gamma  t^2+2\right)-1\right)+\zeta ^4 t^2 \left(\beta  \left(\gamma  t^2-1\right)+2\right)}{\beta  \left(a_0^4 \gamma +2 a_0^2 \gamma  \zeta ^2 t^2+\zeta ^4 t^2 \left(\gamma  t^2+3\right)\right)}.
\end{equation}

\begin{figure}[H]
\centering
\subfigure[]{\includegraphics[width=7.5cm,height=5cm]{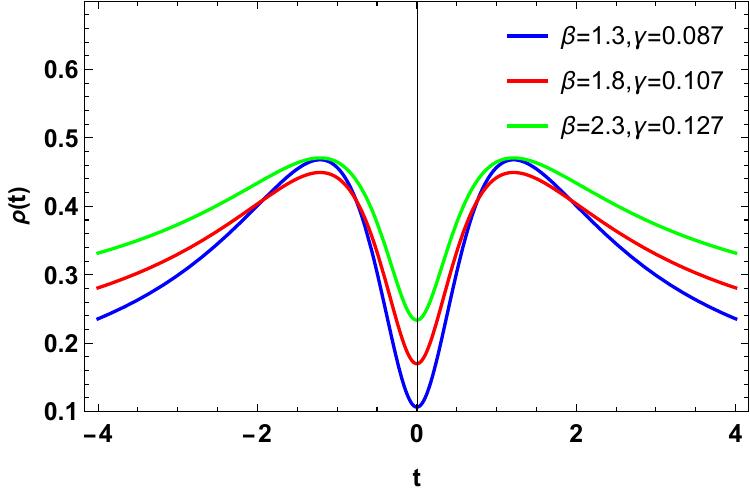}}\,\,\,\,\,\,\,\,\,\,\,\,
\subfigure[]{\includegraphics[width=7.5cm,height=5.15cm]{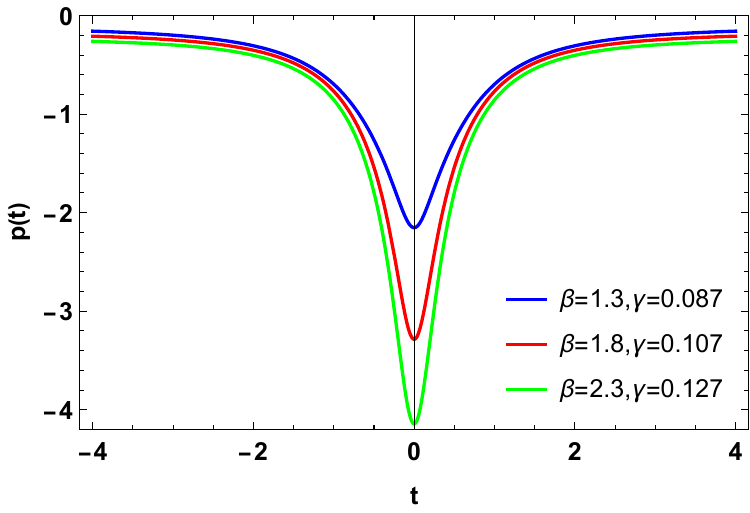}}\,\,\,\,\,\,\,\,\,\,\,\,
\subfigure[]{\includegraphics[width=7.5cm,height=5.2cm]{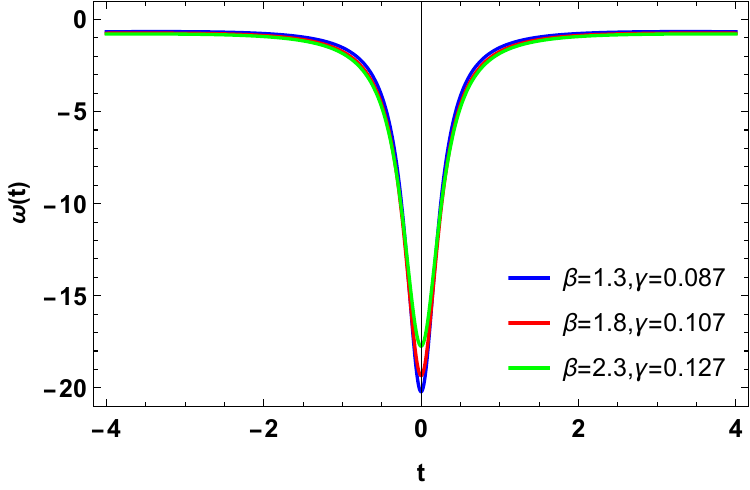}}
\caption{Profile of energy density $\rho(t)$, pressure $p(t)$, and  EoS parameter $\omega(t)$  for $a_0=1$ and  $\zeta=0.824$, with different values of other model parameters $\beta$ and $\gamma$ against cosmic time.}\label{3}
\end{figure}

\justify The obtained expressions for the energy density, pressure, and the EoS parameter possesses bouncing parameter $\zeta$ and the model parameters $\beta$ and $\gamma$. We presented the profile of the above cosmological parameters against cosmic time for the appropriate values of the model parameters as shown in Figure \ref{3} (a), \ref{3} (b), and \ref{3}(c) respectively.

\justify  Figure \ref{3} (a) presents profile of the energy density for the chosen parameter values. In order to satisfy some specific energy conditions, the energy density must be positive through the entire domain of cosmic time . Therefore, we have considered different values of parameters $\beta$ and $\gamma$, with particular value of $a_0$ and $\zeta$, in order to satisfy the positive criteria of the energy density. The energy density satisfies the positivity condition for the range of $\beta \in (0.5,\infty)$, and $\gamma \in(0.05,\infty )$. As the bounce time approaches, the energy density rises dramatically, reaches a peak, and then falls off at the bouncing epoch. Following the bounce, it climbs briefly again before decreasing as cosmic time advances. Figure \ref{3} (b) presents profile of the pressure component against cosmic time. During the cosmic evolution, the pressure was a negative quantity. The pressure component is negative for the range of $\beta \in (0.5,\infty)$, and $\gamma \in(0.015,\infty )$. It exhibits a tiny negative value at the pre-bounce epoch and then increases to a high negative value at the bounce. Further, it approaches to a tiny negative value in the post-bouncing zone. Figure \ref{3} (c) presents profile of the EoS parameter for different values of model parameters $\beta$ and $\gamma$ for model I, where the bouncing parameter has a significant impact on the EoS parameter for non-zero constant values of $a_0$ and $\zeta$. The EoS parameter exhibits negativity for the range of $\beta \in (0.49,\infty)$, and $\gamma \in(0,\infty )$. The EoS parameter corresponding to the model I crosses the $\Lambda$CDM line and lies in the phantom region near the bouncing epoch, whereas it remains in the quintessence phase for the region away from the bouncing epoch. The EoS parameter follows a similar pattern for both the negative and positive time zones.

\justify The cosmological parameters such as energy density $\rho$, pressure $p$ and the EoS parameter $\omega$ for the $f(R,L_m)$ model II, obtained by using equation \eqref{3b} in \eqref{2g} and\eqref{2h} are as follows:


\begin{equation}\label{3d}
\rho(t)=\frac{3a_0^2 \zeta ^4 \left(t^2-12 \lambda \right)+3\zeta ^6 t^2 \left(t^2-36 \lambda \right)}{\alpha  \left(a_0^2+\zeta ^2 t^2\right)^3},
\end{equation}


\begin{equation}\label{3e}
p(t)=-\frac{\zeta ^2 \left(3 a_0^2 \zeta ^2 \left(t^2-12 \lambda \right)+2 a_0^4+\zeta ^4 t^2 \left(36 \lambda +t^2\right)\right)}{\alpha  \left(a_0^2+\zeta ^2 t^2\right)^3},
\end{equation}

\begin{equation}\label{3f}
\omega(t)= -\frac{\zeta ^2 \left(3 a_0^2 \zeta ^2 \left(t^2-12 \lambda \right)+2 a_0^4+\zeta ^4 t^2 \left(36 \lambda +t^2\right)\right)}{3 \left(a_0^2 \zeta ^4 \left(t^2-12 \lambda \right)+\zeta ^6 t^2 \left(t^2-36 \lambda \right)\right)} .
\end{equation}

\begin{figure}[H]
\centering
\subfigure[]{\includegraphics[width=7.5cm,height=5.16cm]{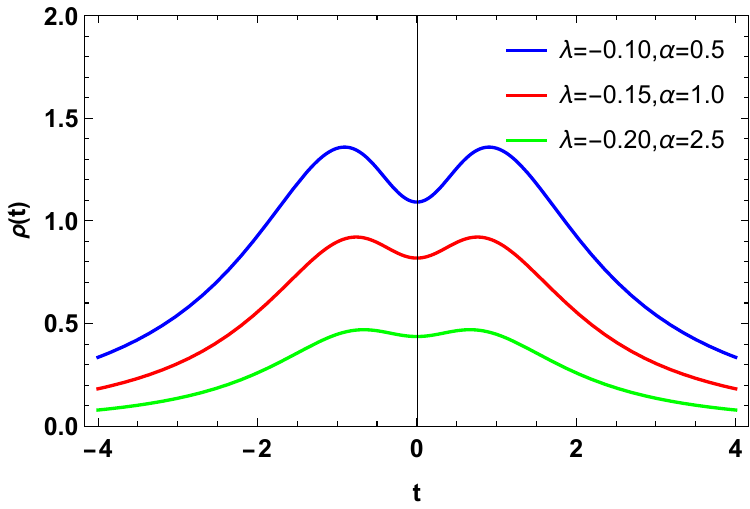}}\,\,\,\,\,\,\,\,\,\,\,\,
\subfigure[]{\includegraphics[width=7.5cm,height=5cm]{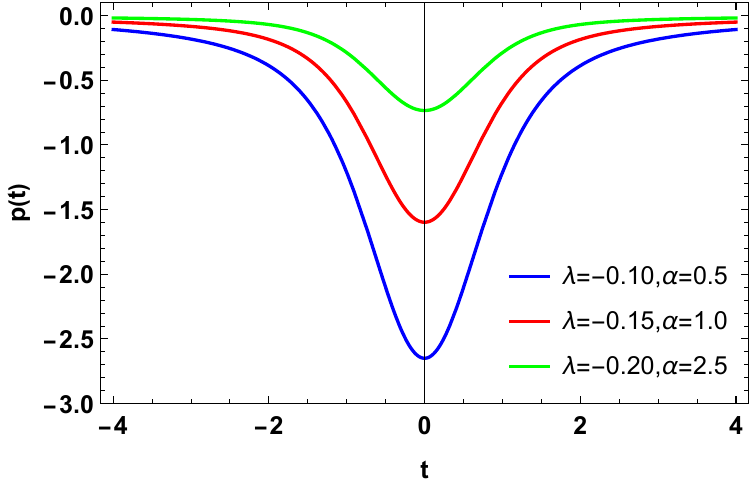 }}\,\,\,\,\,\,\,\,\,\,\,\,
\subfigure[]{\includegraphics[width=7.5cm,height=5.2cm]{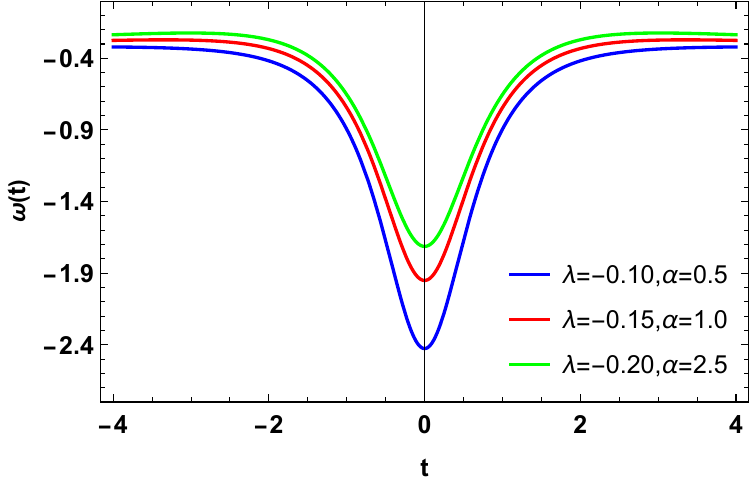}}
\caption{Profile of energy density $\rho(t)$, pressure $p(t)$, and  EoS parameter $\omega(t)$ for $a_0=1$ and  $\zeta=0.624$, with different values of other model parameters $\lambda$ and $\alpha$ against cosmic time.}\label{4}
\end{figure}

\justify The above equations show that energy density, pressure, and EoS parameter are functions of the bouncing parameter $\zeta$ and the model parameters $\alpha$ and $\lambda$. In the present manuscript, we investigated appropriate values of these parameters and plotted them versus cosmic time, as shown in Figure \ref{4} (a), \ref{4} (b), and \ref{4} (c) respectively.

\justify Figure \ref{4} (a) depicts the graphical representation of energy density behaviour for the chosen parameters. The energy density should be positive throughout the span of cosmic history in order to fulfill specific energy conditions. As a result of this, we have taken into account different values of model parameters $\lambda$ and $\alpha$, with perticular value of $a_0$ and $\zeta$, in order to guarantee a positive energy density throughout the cosmic development. The energy density satisfies the positivity condition for the range of $\lambda \in (-\infty,-0.02)$, and $\alpha \in(0,\infty )$. As the bounce time approaches, the energy density rises dramatically, reaches a peak, and then falls off at the bouncing epoch. Following the bounce, it climbs briefly again before decreasing as cosmic time advances. Figure \ref{4} (b) displays a graphical representation of pressure across cosmic time. During the cosmic evolution, the pressure was a negative quantity. The pressure component is negative for the range of $\lambda \in (-0.7,0)$, and $\alpha \in(0,\infty )$. It increases from a tiny negative value in the pre-bounce zone to a high negative value at the bounce. Conversely, it increases from a high negative value to a tiny negative value in the post-bouncing zone. Figure \ref{4} (c) depicts the dynamical behaviour of the EoS parameter for different values of model parameters $\lambda$ and $\alpha$ for Model II, where the bouncing parameter has a significant impact on the EoS parameter for non-zero constant values of $a_0$ and $\zeta$. The EoS parameter exhibits negativity for the range of $\lambda \in (-0.67,0)$, and  $\alpha \in (-\infty,\infty)$.  The model crosses the $\Lambda$CDM line and remains in the quintessence phase as the EoS parameter moves away from the bouncing epoch. On either side of the bouncing period, the EoS parameter develops from the phantom region. Both the negative and positive time zones had a similar pattern of behaviour in the growth of the EoS parameter.

\subsection{Energy Conditions}\label{sbsec2}
\justifying
\justify In the context of general relativity, it is believed that the energy conditions are going to stay essentially positive. A set of linear equations called energy conditions, predict that matter should have a positive energy density. The amplification of energy conditions is the root cause of gravitational collapse, the big bang singularity, and the absence of travelable wormholes. As a result, under the assumption of a perfect fluid matter distribution, the energy conditions \cite{Rayc} that can be determined from standard GR. The model parameters affect every energy condition in the aforementioned expression. Following the DEC, matter moves along time-like or null world trajectories, and the SEC implies that gravity is appealing \cite{Ehle,Noji,Capo3,Mand1}. We obtained the following expressions for the NEC (null energy condition), SEC (strong energy condition), and DEC (dominant  energy condition) corresponding to the $f (R,L_m)$ model I, obtained by using equations \eqref{n6} and \eqref{n7},

\begin{equation}\label{n9}
 \rho+p= \left[\frac{\frac{3 \zeta ^4 t^2}{\left(a_0^2+\zeta ^2 t^2\right)^2}+\gamma }{2 \beta -1}\right]^\frac{1}{\beta} \left[1-\frac{a_0^4 \beta  \gamma +2 a_0^2 \zeta ^2 \left(\beta  \left(\gamma  t^2+2\right)-1\right)+\zeta ^4 t^2 \left(\beta  \left(\gamma  t^2-1\right)+2\right)}{\beta  \left(a_0^4 \gamma +2 a_0^2 \gamma  \zeta ^2 t^2+\zeta ^4 t^2 \left(\gamma  t^2+3\right)\right)}\right]\geq0,
\end{equation}

\begin{equation}\label{n10}
 \rho+3p=  \left[\frac{\frac{3 \zeta ^4 t^2}{\left(a_0^2+\zeta ^2 t^2\right)^2}+\gamma }{2 \beta -1}\right]^\frac{1}{\beta} \left[1-3\frac{a_0^4 \beta  \gamma +2 a_0^2 \zeta ^2 \left(\beta  \left(\gamma  t^2+2\right)-1\right)+\zeta ^4 t^2 \left(\beta  \left(\gamma  t^2-1\right)+2\right)}{\beta  \left(a_0^4 \gamma +2 a_0^2 \gamma  \zeta ^2 t^2+\zeta ^4 t^2 \left(\gamma  t^2+3\right)\right)}\right]\geq0,
\end{equation}

\begin{equation}\label{n11}
\rho-p= \left[\frac{\frac{3 \zeta ^4 t^2}{\left(a_0^2+\zeta ^2 t^2\right)^2}+\gamma }{2 \beta -1}\right]^\frac{1}{\beta} \left[1+\frac{a_0^4 \beta  \gamma +2 a_0^2 \zeta ^2 \left(\beta  \left(\gamma  t^2+2\right)-1\right)+\zeta ^4 t^2 \left(\beta  \left(\gamma  t^2-1\right)+2\right)}{\beta  \left(a_0^4 \gamma +2 a_0^2 \gamma  \zeta ^2 t^2+\zeta ^4 t^2 \left(\gamma  t^2+3\right)\right)}\right]\geq0.
\end{equation}

\begin{figure}[H]
\centering
\subfigure[]{\includegraphics[width=7.5cm,height=5cm]{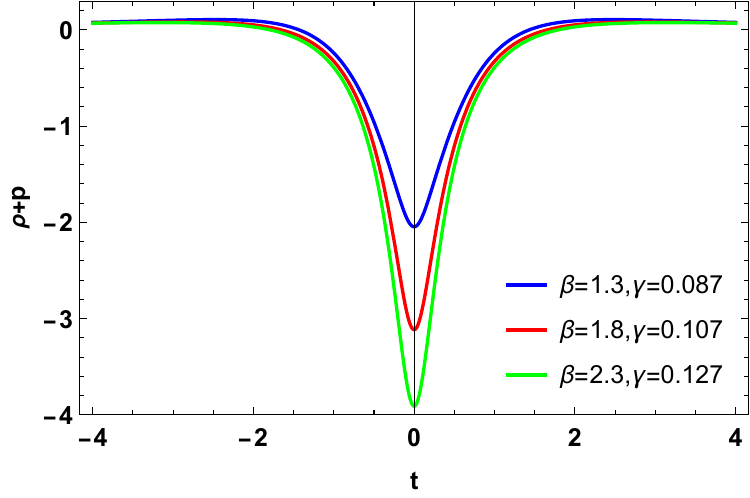}}\,\,\,\,\,\,\,\,\,\,\,\,
\subfigure[]{\includegraphics[width=7.5cm,height=5cm]{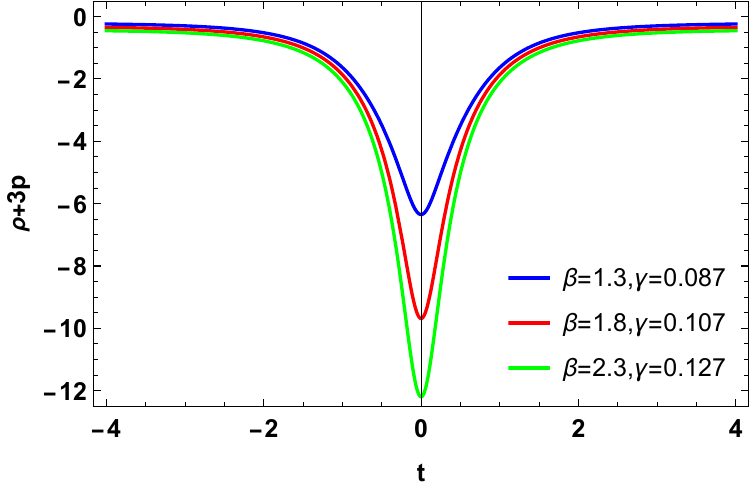}}\,\,\,\,\,\,\,\,\,\,\,\,
\subfigure[]{\includegraphics[width=7.5cm,height=5.2cm]{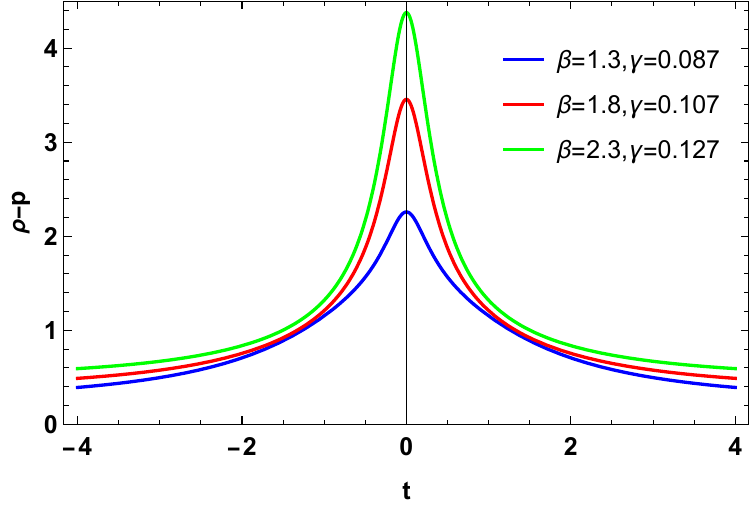}}
\caption{Profile of null energy conditions (NEC), strong energy condition (SEC), and dominant energy conditions (DEC) for $a_0=1$ and $\zeta=0.824$, with different values of other model parameters $\beta$ and $\gamma$ against cosmic time.}\label{5}
\end{figure}

\justify  Figure \ref{5} (a) presents profile of the NEC, whereas Figure \ref{5} (b) presents profile of the SEC, for the chosen parameter values. To complete a non-singular bounce, the EoS parameter must cross the phantom split $\omega <-1$ and then violates the NEC. The bouncing characteristic necessitates the violation of NEC at the bouncing epoch. Figure \ref{5} (a) indicates the clear violation of the NEC, whereas the Figure \ref{5} (b) shows that how the NEC violation leads to the SEC violation, which causes the model to evolve in the phantom phase $\omega <-1$. Additionally, these models shows how the universe is expanding faster than the usual expansion. The figures clearly show that there is no singularity around the bouncing epoch. Figure \ref{5} (c) shows positive behaviour of the DEC and that is expected in case of a perfect fluid type matter distribution. In this case, the NEC and SEC are violated and DEC satiesfies for the range of  $\beta \in (0.59,\infty)$, and $\gamma \in (0,\infty)$.  Moreover, the region close to the bouncing point experiences a uniform shift in energy conditions.

\justify Again, the expressions for the NEC, SEC, and DEC corresponding to the $f (R,L_m)$ model II, obtained by using equations \eqref{3d} and \eqref{3e} reads as,

\begin{equation}\label{4a}
 \rho+p= \frac{2 \zeta ^6 t^2 \left(t^2-72 \lambda \right)-2 a_0^4 \zeta ^2}{\alpha  \left(a_0^2+\zeta ^2 t^2\right)^3}\geq0
 \end{equation}

 \begin{equation}\label{4b}
  \rho+3p=-\frac{6 \zeta ^2 \left(a_0^2 \zeta ^2 \left(t^2-12 \lambda \right)+a_0^4+36 \zeta ^4 \lambda  t^2\right)}{\alpha  \left(a_0^2+\zeta ^2 t^2\right)^3}\geq0
 \end{equation}

 \begin{equation}\label{4c}
 \rho-p= \frac{2 \zeta ^2 \left(a_0^2+2 \zeta ^2 \left(t^2-18 \lambda \right)\right)}{\alpha  \left(a_0^2+\zeta ^2 t^2\right)^2}\geq0
 \end{equation}

\begin{figure}[H]
\centering
\subfigure[]{\includegraphics[width=7.5cm,height=5cm]{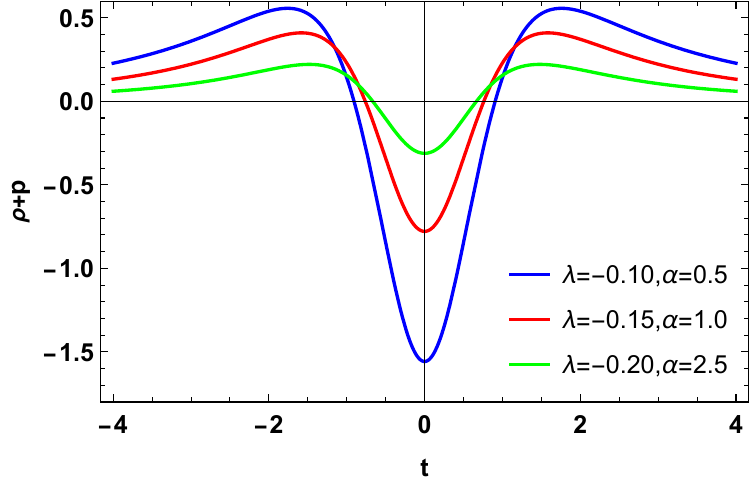}}\,\,\,\,\,\,\,\,\,\,\,\,
\subfigure[]{\includegraphics[width=7.5cm,height=5cm]{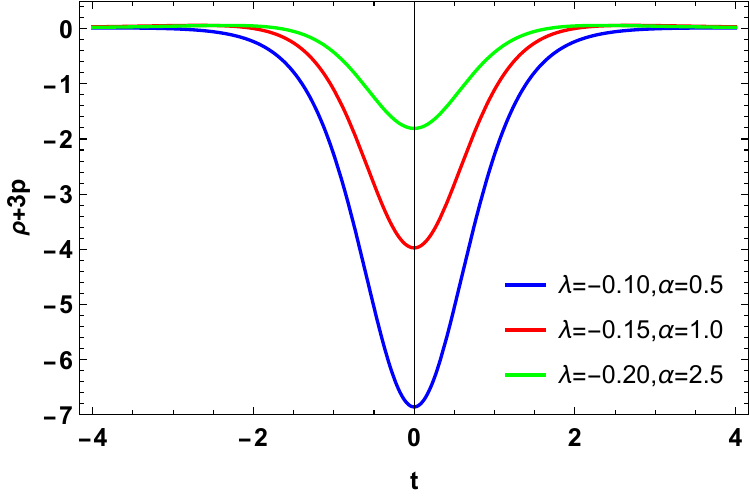}}\,\,\,\,\,\,\,\,\,\,\,\,
\subfigure[]{\includegraphics[width=7.5cm,height=5.2cm]{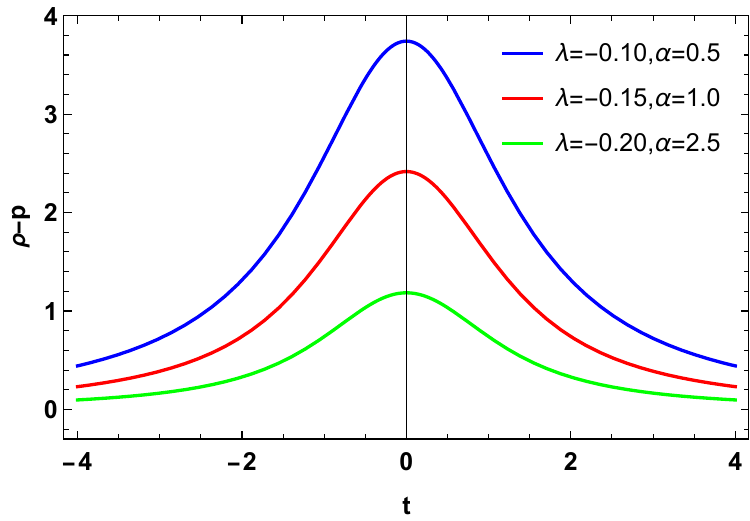}}
\caption{Profile of null energy conditions (NEC), strong energy condition (SEC), and dominant energy conditions (DEC) for $a_0=1$ and $\zeta=0.624$, with different values of other model parameters $\lambda$ and $\alpha$ against cosmic time.}\label{6}
\end{figure}

\justify  Figure \ref{6} (a) depicts the graphical representation of the NEC behaviour for the chosen parameters, whereas Figure \ref{6} (b) depicts the SEC behaviour. Figure \ref{6} (a) illustrates the clear violation of the null energy constraint. Figure \ref{6} (b) shows how the NEC violation leads to the SEC violation, which is nevertheless of less importance in the study. In this case, the NEC is violated for the range of  $\lambda \in (-0.6,\infty)$, and $\alpha \in (0,\infty)$, whereas the SEC is violated for the range of  $\lambda \in (-\infty,0.2)$, and $\alpha \in (0,\infty)$.  The figures clearly show that there is no singularity around the bouncing epoch. However, in the case of a perfect fluid, the DEC is supposed to remain positive, and the same outcome has been seen in Figure \ref{6} (c).  The DEC satisfies for the range of  $\lambda \in (-\infty,0.05)$, and $\alpha \in (0,\infty)$.  Instead, the area close to the bouncing point experiences a uniform shift in energy conditions. The energy conditions NEC and SEC turn negative near the bounce across the positive and negative time zone, in accordance with the predicted matter bounce scenario, and there is a clear indication of violation of the energy condition, which causes the model to evolve in the phantom phase with $\omega <-1$.

\subsection{Stability analysis}\label{sbsec3}
\justifying

\justify The squared speed of sound approach \cite{Suda,Char,Fara} is used in this section to address the stability of bouncing models in $f(R,L_m)$ gravity. The symbol for squared sound speed is $C_s ^2$, and its definition is $C_s ^2=\frac{dp}{d\rho}$. Mechanically stable structures must produce non-negative outcomes when the square of the speed of sound is applied. Therefore, the bouncing models outlined above meet the criteria for being stable for positive values of the squared sound speed $C_s ^2$. The squared speed of sound $C_s ^2$ must remain within zero and one for mechanical stability evaluation. The following is the cosmic time-based equation for the squared speed of sound $C_s ^2$ can be determined,

\justify Corresponding to the model I, we have by using equations \eqref{n6} and \eqref{n7} as
    
\begin{multline}\label{n12}
 C_s^2 =\frac{3 a_0^6 \beta  (4 \beta -3) \gamma +a_0^4 \zeta ^2 \left(6 \beta  (2 \beta -3)+\beta  (20 \beta -13) \gamma  t^2+6\right)+a_0^2 \zeta ^4 t^2 \left(3 \beta  (4 \beta +3)+\beta  (4 \beta +1) \gamma  t^2-12\right)}{3 \beta  (a_0^2-\zeta ^2 t^2) \left(a_0^4 \gamma +2 a_0^2 \gamma  \zeta ^2 t^2+\zeta ^4 t^2 \left(\gamma  t^2+3\right)\right)}+\\ \frac{ \zeta ^6 t^4 \left(-3 \beta +\beta  (5-4 \beta ) \gamma  t^2+6\right)}{3 \beta  (a_0^2-\zeta ^2 t^2) \left(a_0^4 \gamma +2 a_0^2 \gamma  \zeta ^2 t^2+\zeta ^4 t^2 \left(\gamma  t^2+3\right)\right)}.
\end{multline}

\justify Corresponding to the model II, we have by using equations \eqref{3d} and \eqref{3e} as

\begin{equation}\label{6a}
 C_s ^2 =\frac{6 \zeta ^4 t \left(3 a_0^2 \zeta ^2 \left(t^2-12 \lambda \right)+2 a_0^4+\zeta ^4 t^2 \left(36 \lambda +t^2\right)\right) - \zeta ^2 \left(6 a_0^2 \zeta ^2 t+2 \zeta ^4 t \left(36 \lambda +t^2\right)+2 \zeta ^4 t^3\right)}{3 \left(2 a_0^2 \zeta ^4 t+2 \zeta ^6 t \left(t^2-36 \lambda \right)+2 \zeta ^6 t^3\right)-18 \zeta ^2 t \left(a_0^2 \zeta ^4 \left(t^2-12 \lambda \right)+\zeta ^6 t^2 \left(t^2-36 \lambda \right)\right)}.
\end{equation}

\begin{figure}[H]
\centering
\subfigure[]{\includegraphics[width=7.5cm,height=5.2cm]{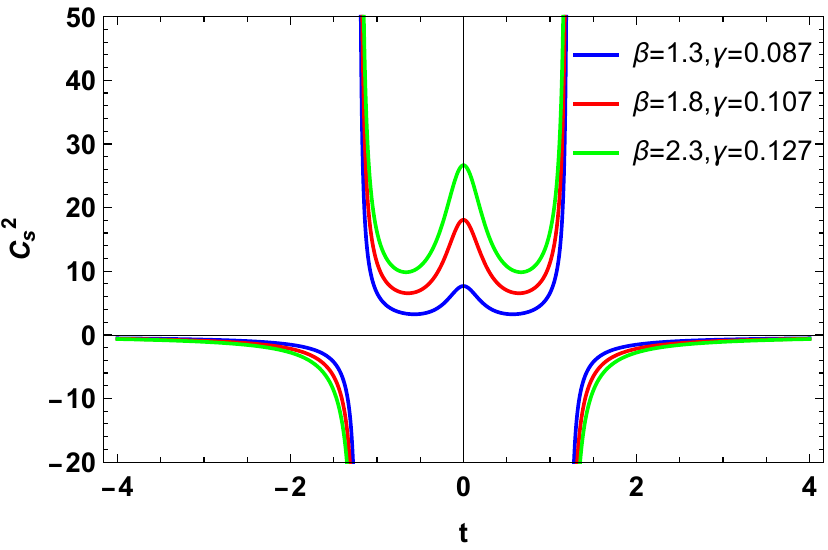}}\,\,\,\,\,\,\,\,\,\,\,\,
\subfigure[]{\includegraphics[width=7.5cm,height=5.2cm]{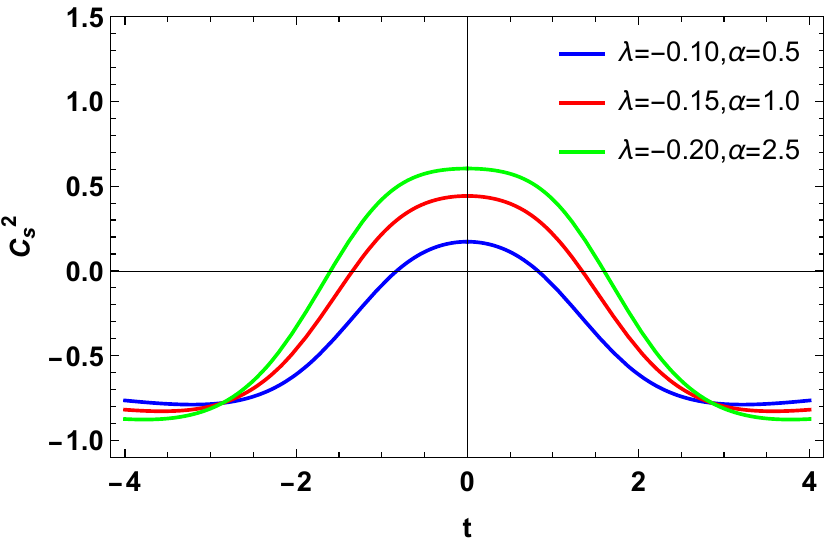 }}
\caption{Profile of stability analysis for $a_0=1$, $\zeta=0.824$ for model I, and $\zeta=0.624$ for model II , with different values of other model parameters $\beta$,$\gamma$ and $\lambda$, $\alpha$ against cosmic time respectively.}\label{7}
\end{figure}

\justify Figure \ref{7} (a) and (b) shows stability analysis for model I and model II. It can be seen from \ref{7} (a) for different values of model parameters $\beta$ and $\gamma$ with bouncing parameter $\zeta$, squared speed of sound shows positive values but still greater than one which predicts unstable behaviour of model I. As demonstrated in Figure \ref{7} (b), the squared speed of sound exhibits positive values between zero and one at a specific time range for various deals of the model parameters $\lambda \in (-0.6,0)$, and  $\alpha \in (-\infty,\infty) $, which predicts stable behaviour for model II. Since the suggested model exhibits bouncing behaviour at the epoch $t = 0$, and the stability range includes the bouncing epoch, we infer the model I is unstable, and the model II is stable.

\section{Conclusions}\label{sec6}
\justifying

\justify In recent years, the singularity problem and the inflationary problem have posed significant challenges for researchers looking to understand the origin and evolution of the universe. Without sufficient observational data, cosmologists have turned to a potential solution known as bouncing cosmology. Bouncing cosmology represents an alternative view of the early universe and offers a solution to the common singularity problem in standard cosmological models. We have investigated the possibility of reproducing specific bouncing models within the $f(R, L_m)$ gravity framework. We have observed the bounce at the epoch $t = 0$ by observing how the dynamical parameters behave. It is confirmed that a cosmic bounce occurs when the Hubble parameter reaches $H = 0$. Another requirement for the bouncing scenario is a violation of NEC, we have confirmed this violation in the range near the bounce. The EoS parameter and energy density behaviour support the bouncing pattern and remain totally in the negative and positive regions, respectively. The geometrical parameters, which include the Hubble parameter and the deceleration parameter, are produced by constructing a parametric modeling of the scale factor of the form $a(t)=(a_0^2 +\zeta^2 t^2)^\frac{1}{2}$. Similar to other cosmographic parameters like the jerk, snap, and lerk parameters, the action of the deceleration parameter during the bounce exhibits singularity. The intriguing aspect of the bouncing model is that the lerk parameter achieves singularity in its positive profile, while the jerk and snap parameters reach singularity at their negative profiles. 

In this work, we attempted to explore the behavior of the bouncing solution in the novel $f(R, L_m)$ gravity. We consider two toy cosmological models of the theory and found the constraint on model parameters to exhibit the bouncing behavior. We consider two non-linear $f(R, L_m)$ cosmological models, particularly, $f(R,L_m)=\frac{R}{2}+\alpha L_m+\lambda R^2$, and $f(R,L_m)=\frac{R}{2}+L_m ^\beta +\gamma$ under the flat FLRW background with perfect fluid type matter distribution. Both the considered $f(R, L_m)$ lagrangian have great significance, providing the impact of corrections to both geometrical part as well as the matter sector respectively. We derive the expressions for the density, pressure, EoS parameter, and different energy conditions, presenting a descriptive picture of the initial circumstances of the universe at the bounce for both considered non-linear models. At the pre-bouncing epoch, the energy density increases, attains a peak, and then further dramatically decreases. At the post-bouncing epoch, it further increases, attains a peak, and then decreases as cosmic time moves forward (see left panels of Figures \ref{3} and \ref{4}). Corresponding to both the assumed non-linear models, the pressure components exhibit high negative values near the bounce, whereas it exhibits a tiny negative value away the bouncing epoch (see right panels of Figures \ref{3} and \ref{4}). Due to the non-vanishing nature of the scale factor, we discovered the beginning conditions of density and pressure in the universe are finite, eliminating the first singularity problem. For both the models, we found that, in the region near the bounce, the EoS parameter crosses the $\Lambda$CDM line ($\omega=-1$) and lies in the phantom phase ($\omega<-1$), whereas in the region far away from the bouncing epoch, it lies in the quintessence phase ($\omega>-1$) (see lower panels of Figures \ref{3} and \ref{4}). In order to achieve any non-singular bounce with a standard matter source, it is required to demonstrate the violation of NEC and SEC at the bounce region. The violation of NEC and SEC near the bouncing epoch have been presented in Figure \ref{5} for model I, and in Figure \ref{6} for model II. However, the DEC shows positive behaviour for both models. The fact that the SEC has been broken repeatedly during the course of cosmic history is interesting because it suggests that the expansion of the universe accelerated in the late cosmic period. Finally, using the squared speed of sound ($C_s ^2$) method, we looked into the stability of the obtained bouncing solutions, presented in  Figure \ref{7}. We found that model I shows unstable behaviour, whereas model II shows stable behaviour.  

\section*{Data Availability Statement}

There are no new data associated with this article.

\section*{Acknowledgments} \label{sec10}
L.V.J. acknowledges UGC, Govt. of India, New Delhi, for awarding JRF (NTA Ref. No.: 191620024300). R.S. acknowledges UGC, New Delhi, India for providing Senior Research Fellowship (UGC-Ref. No.: 191620096030). PKS acknowledges Science and Engineering Research Board, Department of Science and Technology, Government of India for financial support to carry out Research project No.: CRG/2022/001847 and IUCAA, Pune, India for providing support through the visiting Associateship program. We are very much grateful to the honorable referees and to the editor for the illuminating suggestions that have significantly improved our work in terms of research quality, and presentation.

\end{document}